\documentclass[iop,numberedappendix, appendixfloats]{emulateapj}

\usepackage{times,natbib,graphicx,amsmath}

\newcommand{\nustar}{\textit{NuSTAR}}

\newcommand{\nicer}{\textit{NICER}}
\newcommand{\xmm}{{\it XMM-Newton}}
\newcommand{\integral}{{\it INTEGRAL}}
\newcommand{\beppo}{{\it BeppoSAX}}

\newcommand{\mdot}{$\dot{\mathrm{m}}$}
\newcommand{\fluxcgs}{ergs~s$^{-1}$~cm$^{-2}$}
\newcommand{\lumcgs}{ergs~s$^{-1}$}
\newcommand{\rin}{$R_{in}$}
\newcommand{\risco}{$R_{\mathrm{ISCO}}$}

\newcommand{\cps}{counts s$^{-1}$}

\shorttitle{{\it NuSTAR} observations of accreting atolls}
\shortauthors{Ludlam et al.}

\begin{document}

\title{\nustar\ Observations of the Accreting Atolls GX 3+1, 4U 1702$-$429, 4U 0614+091, and 4U 1746$-$371}
\author{R. M. Ludlam\altaffilmark{1},
J. M. Miller\altaffilmark{1},
D. Barret\altaffilmark{2,3},
E. M. Cackett\altaffilmark{4},
B. M. Coughenour\altaffilmark{4},
T. Dauser\altaffilmark{5},
N. Degenaar\altaffilmark{6},
J. A. Garc\'{i}a\altaffilmark{5,7}, 
F. A. Harrison\altaffilmark{7},
F. Paerels\altaffilmark{8}
}
\altaffiltext{1}{Department of Astronomy, University of Michigan, 1085 South University Ave, Ann Arbor, MI 48109-1107, USA}
\altaffiltext{2}{Universit de Toulouse; UPS-OMP; IRAP; Toulouse, France}
\altaffiltext{3}{CNRS; Institut de Recherche en Astrophysique et Plantologie; 9 Av. colonel Roche, BP 44346, F-31028 Toulouse cedex 4, France}
\altaffiltext{4}{Department of Physics \& Astronomy, Wayne State University, 666 W. Hancock St., Detroit, MI 48201, USA}
\altaffiltext{5}{Remeis Observatory \& ECAP, Universit\"{a}t Erlangen-N\"{u}rnberg, Sternwartstr. 7, 96049, Bamberg, Germany}
\altaffiltext{6}{Anton Pannekoek Institute for Astronomy, University of Amsterdam, Pastbus 94249, 1090 GE Amsterdam, The Netherlands}
\altaffiltext{7}{Cahill Center for Astronomy and Astrophysics, California Institute of Technology, Pasadena, CA 91125}
\altaffiltext{8}{Columbia Astrophysics Laboratory, 550 West 120th Street, New York, NY 10027, USA}

\begin{abstract} 
Atoll sources are accreting neutron star (NS) low-mass X-ray binaries. We present a spectral analysis of four persistent atoll sources (\mbox{GX 3+1}, \mbox{4U~1702$-$429}, \mbox{4U~0614+091}, and \mbox{4U 1746$-$371}) observed for $\sim20$ ks each with \nustar\ to determine the extent of the inner accretion disk. These sources range from an apparent luminosity of $0.006-0.11$ of the Eddington limit (assuming the empirical limit of $3.8\times10^{38}$ \lumcgs). Broad Fe emission features shaped by Doppler and relativistic effects close to the NS were firmly detected in three of these sources. The position of the disk appears to be close to the innermost stable circular orbit (ISCO) in each case. For \mbox{GX 3+1}, we determine $R_{in}=1.8^{+0.2}_{-0.6}\ R_{\mathrm{ISCO}}$ (90\% confidence level) and an inclination of $27^{\circ}-31^{\circ}$. For \mbox{4U~1702$-$429}, we find a $R_{in}=1.5_{-0.4}^{+1.6}\ R_{\mathrm{ISCO}}$ and inclination of $53^{\circ}-64^{\circ}$. For \mbox{4U~0614+091}, the disk has a position of $R_{in}=1.3_{-0.2}^{+5.4}\ R_{\mathrm{ISCO}}$ and inclination of $50^{\circ}-62^{\circ}$. If the disk does not extend to the innermost stable circular orbit, we can place conservative limits on the magnetic field strength in these systems in the event that the disk is truncated at the Alfv\'{e}n radius. This provides the limit at the poles of $B\leq6.7\times10^{8}$ G, $3.3\times10^{8}$ G, and $14.5\times10^{8}$ G for \mbox{GX 3+1}, \mbox{4U~1702$-$429}, and \mbox{4U~0614+091}, respectively. For \mbox{4U 1746$-$371}, we argue that the most plausible explanation for the lack of reflection features is a combination of source geometry and strong Comptonization. We place these sources among the larger sample of NSs that have been observed with \nustar. 
\end{abstract}


\section{Introduction}

Accretion onto compact objects in low-mass X-ray binaries (LMXBs) occurs via a disk that formed through Roche-lobe overflow from the envelope of the roughly stellar mass companion star. Persistently accreting neutron star (NS) LMXBs are separated into two categories: \lq \lq Z" and \lq \lq atoll". These classifications derive their name from the shape they trace out in color-color and hardness-intensity diagrams \citep{hvp89}. Z sources are very luminous as they likely accrete at near Eddington luminosities ($0.5-1.0$ L$_{\mathrm{Edd}}$: \citealt{van05}), whereas atoll sources are typically less luminous ($\sim0.001-0.5$ L$_{\mathrm{Edd}}$). Transient systems that alternate between periods of active accretion and quiescence often exhibit atoll-like or Z-like behavior during outburst. Some sources are even able to transition between the two classes (e.g., XTE J1701$-$462, \citealt{homan10}) suggesting a trend with average mass accretion rate.

Atoll sources generally have two spectral states analogous to black hole (BH) LMXBs: 1) a hard state characterized by power-law emission with little thermal emission, and 2) a soft state dominated by thermal emission. Additionally, they show intermediate behavior as the sources transition between these states (see \citealt{wijnands17} for a recent discussion on the detailed morphology). 

\citet{lin07} analyzed the spectrum of two atoll type transients (Aquila X-1 and 4U 1608$-$52) to devise a ``hybrid" model for the hard and soft spectral states. The hard state can be described by a single-temperature blackbody to account for boundary layer emission (where material from the disk reaches the surface of the NS) and a power-law component to account for Comptonization. The soft state can be described by a double thermal model comprised of a multi-temperature blackbody for disk emission and a single-temperature blackbody with the addition of a power-law component. This provided a coherent picture of the spectral evolution (e.g., the thermal components follow $L_{\mathrm{X}}\propto T^{4}$) and timing behavior of these sources that is analogous to BHs, which has been utilized for a number of other NS LMXBs (e.g., \citealt{cackett08}, \citeyear{cackett09}, \citeyear{cackett10}; \citealt{lin10}; \citealt{miller13}; \citealt{chiang16}).  

The hard X-ray photons that originate from the boundary layer or coronal region can illuminate the disk and be reprocessed by the material therein before being re-emitted. This reprocessed emission is known as the \lq \lq reflection" spectrum. Doppler and relativistic effects are imprinted on features in the reflection spectrum, such as Fe K$_{\alpha}$, yielding information about accretion flow \citep{fabian89}. The strength of these effects increase with proximity to the compact object, thus allowing the position of the inner accretion disk to be determined from the shape of the Fe line profile. The accretion disk around a NS has to truncate at or prior to the surface, hence reflection studies in NS LMXBs can provide upper limits on the radial extent of these objects (\citealt{cackett08}; \citealt{miller13}; \citealt{ludlam17a}) or indicate the presence of a boundary layer or strong magnetic field (\citealt{cackett09}; \citealt{papitto09}; \citealt{king16}; \citealt{ludlam17b}, \citeyear{ludlam17c}; \citealt{vandeneijnden17}). 

The location of the inner edge of the accretion disk around compact objects is thought to be dependent on the mass accretion rate, \mdot, of the system (\citealt{gonzalez14} for a review). Indeed, the overall spectral evolution of atolls does change with \mdot\ \citep{gladstone07}, but there does not appear to be a clear one-to-one correlation with the inner disk position (\citealt{cackett10}; \citealt{chiang16}; \citealt{ludlam17a}). When looking at a sample of NS LMXBs from which the inner disk radius could be determined via reflection, some atoll sources were consistent with the innermost stable circular orbit (ISCO) at as low as 1\% L$_{\mathrm{Edd}}$, whereas others were truncated at higher accretion rates near 10\% L$_{\mathrm{Edd}}$ \citep{ludlam17a}. This suggests a more complex behavior for the position on the inner accretion disk that relies on more than just \mdot\ (e.g., the importance of the magnetosphere or boundary layer).

We present a sample of four persistently accreting atoll sources that were approved an initial $\sim20$ ks observation each with \nustar\ \citep{harrison13} during GO Cycle 3: \mbox{GX 3+1}, \mbox{4U~1702$-$429}, \mbox{4U~0614+091}, and \mbox{4U 1746$-$371}. We search for the presence of reflection features and place constraints on the position of the inner disk. \nustar\ has been an exceptional tool for reflection studies due to its large energy bandpass from $3-79$ keV, as well as its high effective collecting area that is free from instrumental effects such as pile-up.
Our sample spans the range of 0.006 to 0.11 L$_{\mathrm{Edd}}$, assuming the empirical Eddington limit of $3.8\times10^{38}$ \lumcgs \citep{kuulkers03}. The paper is structured in the following format: the subsequent subsections provide background on each source (\S 1.1, 1.2, 1.3, 1.4), \S 2 presents the observations and data reduction, \S 3 discusses the spectral analysis and results, \S 4 provides a discussion of the results, \S 5 summarizes the discussion. 

\subsection{GX 3+1}
\mbox{GX 3+1} is known to exhibit both Type-I X-ray bursts (\citealt{makishima83}; \citealt{kuulkers00}; \citealt{chenevez06}) and longer duration superbursts from carbon burning \citep{kuulkers02}. Imposing the assumption that the Type-I X-ray bursts are Eddington limited provides a maximum distance to the source of 6.5 kpc \citep{galloway08}. A broad Fe line was first detected in this source with {\it Beppo}SAX \citep{oosterbroek01}. {\it XMM-Newton} observations of the source confirmed the presence of a relativistically shaped Fe line and revealed potential lower-energy features due to Ar {\sc xviii} and Ca {\sc xix} \citep{piraino12}. The Fe line profile suggested that the inner disk was located at a distance of $\sim25\ R_{g}$ (where $R_{g}=GM/c^{2}$) with an inclination of $35^{\circ}-44^{\circ}$ when fit with a simple {\sc diskline} model \citep{fabian89}. \citet{pintore15} analyzed {\it XMM-Newton} and {\it INTEGRAL} observations from 2010 which suggested that the disk was closer to the NS at $\sim10\ R_{g}$ with an inclination of $\sim35^{\circ}$ when accounting for the entire reflection spectrum.

\subsection{4U 1702$-$429}
\mbox{4U~1702$-$429} is a burster \citep{swank76} located at a maximum distance of 5.65 kpc \citep{galloway08}, assuming that the Type-I X-ray bursts are Eddington limited. Burst oscillations were detected at a frequency of 330 Hz with the {\it Rossi X-ray Timing Explorer} ({\it RXTE}) \citep{markwardt99}, which is indicative of the spin frequency of the NS. This spin frequency corresponds to a dimensionless spin parameter ($a=cJ/GM^{2}$) of 0.155 \citep{braje00}, assuming a NS mass of 1.4 $M_{\odot}$, a 10 km radius, and softish equation of state of the \lq FPS' model \citep{cook94}. 
\citet{iaria16}  provided the first investigation of the broadband spectrum using observations from both {\it XMM-Newton} and {\it INTEGRAL}. This revealed a broad Fe line component that originated from a significantly truncated disk with an inner disk radius of $31_{-12}^{+25}\ R_{g}$ and implied an inclination of $44_{-6}^{+33}$$^{\circ}$.

\subsection{4U 0614+091}
\mbox{4U~0614+091} is an ultracompact X-ray binary (UCXB) with an orbital period of $\sim50$ minutes \citep{shahbaz08} located at a distance of 3.2 kpc \citep{kuulkers10}. Type-I X-ray bursts have been detected in this system (\citealt{swank78}; \citealt{brandt92}) confirming that the compact object is a NS. 
A spin frequency of 415 Hz \citep{strohmayer08} was determined from burst oscillations, which translates to $a=0.2$ using the formalism from \citet{braje00}. The companion star is either a CO or ONe white dwarf due to carbon and oxygen emission lines present in the optical spectrum \citep{nelemans04}. \citet{madej10} detected relativistically broadened O {\sc viii} Ly$_{\alpha}$ emission in the {\it XMM-Newton} Reflection Grating Spectrometer detectors. When modeling the O {\sc viii} Ly$_{\alpha}$ emission with a relativistic line profile, this emission appeared to originate from the innermost region ($\sim6\ R_{g}$) of a highly inclined disk ($i\approx54^{\circ}$). 
A follow up study using the \xmm/EPIC-pn data additionally revealed the presence of an Fe line feature \citep{madej14}. Spectral modeling of both features via a specially tailored reflection spectrum model for an overabundance of C and O, {\sc xillver$_{\mathrm{CO}}$}, supported the previous results of a disk close to the ISCO with a moderate inclination. 

\subsection{4U 1746$-$371}
\mbox{4U 1746$-$371} is associated with the globular cluster NGC 6441. Variable stars in the cluster establish the distance to NGC 6441 to between $10.4-11.9$ kpc \citep{pritzl01}. The source is known to experience Type-I X-ray bursts (\citealt{sztajno87}) and is a dipping source \citep{parmar89}, which indicates that the system is highly inclined. The periodicity of the dipping observed with {\it Ginga} suggested an orbital period of $\sim5.7$ hours \citep{sansom93}, but follow-up studies with {\it RXTE} found a smaller period of $5.16\pm0.01$ hours \citep{bcs04}. {\it Hubble Space Telescope} imaging of the globular cluster identified five possible optical counterparts, but the field was too dense to isolate a single counterpart as the companion to \mbox{4U 1746$-$371} \citep{deutsch98}. \citet{asai00} found evidence of a potential narrow Fe line in {\it ASCA} data, whereas a later investigation by \citet{diaztrigo06} with \xmm\ suggested a broad Fe emission component (modeled with a Gaussian at 6.4 keV that provided a 3.5$\sigma$ improvement) with a potential narrow Fe {\sc xxvi} absorption line. 

\begin{table*}[t!]
\caption{\nustar\ Observation Information}
\label{tab:obs} 
\begin{center}
\begin{tabular}{lcccccc}
\hline
Source & Mission &  Obs ID  & Date (yyyy-mm-dd) & Exp (ks) & Net Rate (cts s$^{-1}$)\\
\hline
GX 3+1 & \nustar & 30363001002  & 2017-10-17 & 13.7 &  147.9\\
4U 1702$-$429 & \nustar & 30363005002 &2017-08-29& 21.7 & 97.1 \\
4U 0614+091 & \nustar & 30363002002  & 2017-12-01& 15.1 & 28.3\\
4U 1746$-$371 & \nustar& 30363004002  & 2018-02-10 &19.8 & 10.7\\
\hline
\end{tabular}
\end{center}
\end{table*}

\setcounter{footnote}{0}

\section{Observations and Data Reduction}

Table \ref{tab:obs} provides the ObsIDs, date, exposure time, and net count rate of each \nustar\ observation. 
We followed the standard data reduction procedures using {\sc nustardas} v1.8.0 with {\sc caldb} 20180126 for each observation. \mbox{GX 3+1} and \mbox{4U~0614+091} required additional parameters to be called when utilizing the {\sc nupipeline} tool. \mbox{GX 3+1} is considered a formally bright source for \nustar\ ($>100$ \cps), so we applied {\sc statusexpr}=\lq\lq STATUS==b0000xxx00xxxx000" to correct for high count rates. The observation of \mbox{4U~0614+091} occurred during high intervals of background near the South Atlantic Anomaly (SAA), therefore we imposed {\sc saacalc} $=$ 3, {\sc saamode}$=$strict, and {\sc tentacle}$=$yes to reduce these periods. Using the {\sc nuproducts} tool we created lightcurves and spectra for each observation using a circular extraction region with a radius of 100$^{\prime \prime}$ centered around the source to produce a source spectrum for both the Focal Plane Modules (FPMA and FPMB). We use another 100$^{\prime \prime}$ radial region away from the source for the purpose of background subtraction. No Type-I X-ray bursts were present in the lightcurves. 
Although it is not generally recommended to combine data from the FPMA/B since the introduced systematics may be larger than the statistical errors in high signal-to-noise observations, most of the sources in our sample have a low signal-to-noise. Therefore, we combine the FPMA/B using {\sc addascaspec} to improve the signal-to-noise and provide a uniform analysis among sources in this sample.
The data were binned by a factor of 3 using {\sc grppha} \citep{choudhury17}.

\section{Spectral Analysis and Results}

We use {\sc xspec} version 12.9.1m \citep{arnaud96} to perform our spectral analysis. Errors are generated from Monte Carlo Markov Chains of length 500,000 in order to simultaneously probe the entire $\chi^{2}$ parameter space and quoted at the 90\% confidence level. We use {\sc tbabs}  \citep{wilms} to account for absorption along the line of sight to each source with the abundance set to {\sc wilm} \citep{wilms} and {\sc vern} cross-sections \citep{vern}. The upper limit on the energy range in which each source is modeled is background limited. 

Since the \nustar\ low-energy bandpass cuts off at 3 keV, we are unable to constrain the absorption column along the line of sight. We utilize archival \xmm/RGS data to fit absorption edges with {\sc tbnew}\footnote{http://pulsar.sternwarte.uni-erlangen.de/wilms/research/tbabs/} to determine N$_{\mathrm{H}}$ for each source. Note that these observations are not simultaneous with our \nustar\ observations. The RGS data were fit in the $0.45-2.1$ keV energy range. We find N$_{\mathrm{H}}=2.42\pm0.02\times10^{22}$ cm$^{-2}$ for GX~3+1, N$_{\mathrm{H}}=2.32\pm0.06\times10^{22}$ cm$^{-2}$ for 4U~1702$-$429,  N$_{\mathrm{H}}=3.46\pm0.01\times10^{21}$ cm$^{-2}$ for 4U~0614+091, and  N$_{\mathrm{H}}=3.76\pm0.01\times10^{21}$ cm$^{-2}$ for 4U~1746$-$371. We compare these values to those reported in previous studies on these sources and find good agreement (e.g., GX~3+1: \citealt{pintore15}, 4U~1702$-$429: \citealt{mazzola18}, 4U~0614+091: \citealt{madej10}, \citeyear{madej14}, 4U~1746$-$371: \citealt{diaztrigo06}). We therefore fix N$_{\mathrm{H}}$ in the following \nustar\ fits since column density is dominated by the interstellar medium and does not vary with spectral state \citep{miller09}. This is certainly true for low inclination source, but there can be deviations for high inclination sources due to obscuration within the system itself during dips or eclipses. However, note that the higher inclination sources in this sample do not show dips during our observations.

The continuum is modeled based upon the prescription in \citet{lin07} for atoll sources. This is largely motivated by the existence of self-consistent reflection models based upon these components. We check, when appropriate, that our choice of continuum does not bias our results.  Only \mbox{GX 3+1} showed a very soft spectrum consistent with the soft state. The spectrum was modeled with a multi-temperature blackbody ({\sc diskbb}; \citealt{mitsuda84}), single-temperature blackbody ({\sc bbody}), and power-law component ({\sc powerlaw}). The other sources exhibited spectra consistent with hard state and are therefore modeled with a power-law component ({\sc cutoffpl}) and single-temperature blackbody ({\sc bbody}) when needed.

When reflection features are present in the spectra, we utilize the self-consistent reflection model {\sc relxill} \citep{garcia14} or the preliminarily flavor of this model tailored to thermal emission from a NS {\sc relxillns} (Dauser, Garc\'{i}a, \& Ludlam {\it in prep.}). The major difference between these models is the illuminating continuum that provides the hard X-rays that shape the resulting reflection spectrum. {\sc relxill} uses  a cutoff power-law input spectrum whereas {\sc relxillns} assumes a blackbody is irradiating the disk. The model components of {\sc relxill} are as follows: $q_{1}$ is the inner emissivity index, $q_{2}$ is the outer emissivity index, $R_{break}$ is the break radius between the two emissivity indices, $a$ is the dimensionless spin parameter, $i$ is the inclination in degrees, $R_{in}$ is the inner disk radius in units of the innermost stable circular orbit ($R_{\mathrm{ISCO}}$), $R_{out}$ is the outer disk radius in units of gravitational radius ($R_{g}=GM/c^{2}$), $\Gamma$ is the photon index of the input cutoff power-law, $\log \xi$ is the log of the ionization parameter, $A_{Fe}$ is the iron abundance of the system normalized to the Sun, E$_{\mathrm{cut}}$ is the cutoff energy, $f_{\mathrm{refl}}$ is the reflection fraction, $z$ is the redshift to the object, and norm represents the normalization of the model. {\sc relxillns} has similar parameters with the addition of $\log n$ (cm$^{-3}$) to vary the density of the accretion disk and $kT_{bb}$ (keV) for the input blackbody spectrum instead of $\Gamma$ and E$_{\mathrm{cut}}$.

When using the reflection models we impose the following: a single emissivity index $q= q_{1}=q_{2}$, a redshift of $z=0$ since these are Galactic sources, a spin of $a=0$ since most NS in LMXBs have $a\leq0.3$ (\citealt{galloway08}; \citealt{miller11}) and the ISCO is a slowly varying function in this regime, and a large outer disk radius of $R_{out}=990\ R_{g}$. For reference, $1\ R_{\mathrm{ISCO}}=6\ R_{g}$ for $a=0$ \citep{bardeen72}. In the case of \mbox{4U~1702$-$429} and \mbox{4U~0614+091} where burst oscillations imply $a=0.155$ and $a=0.199$, the assumption of $a=0$ is a marginal difference of $<0.7\ R_{g}$ for the position of the ISCO. \\

\begin{figure*} 
\begin{center}
\includegraphics[width=\textwidth]{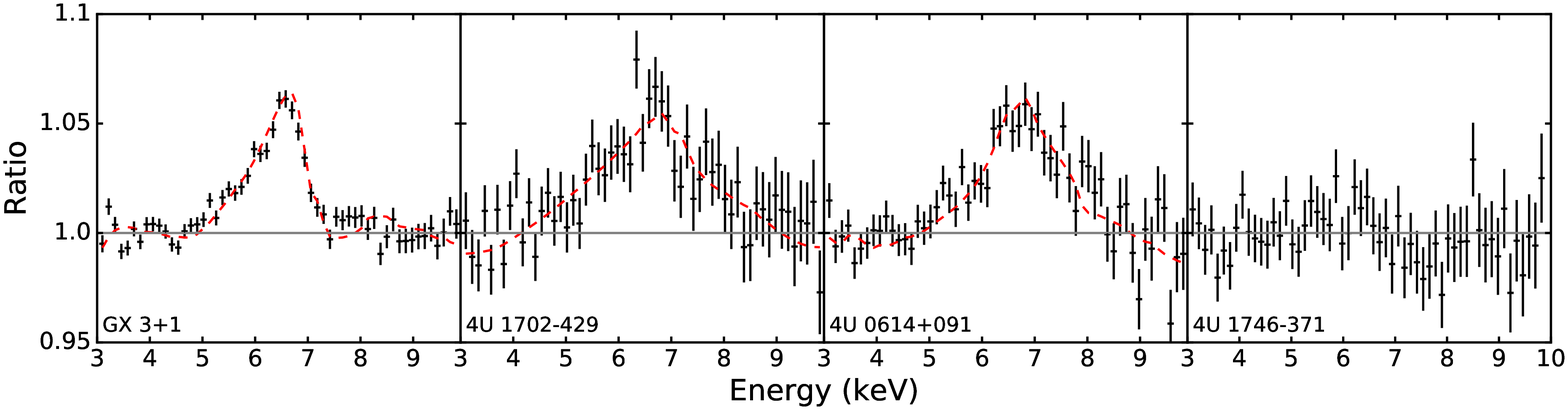}
\caption{Ratio of the data to the continuum model for the atoll sources. The $5.0-8.0$ keV energy band was ignored to prevent the Fe line region from skewing the fit. The dashed red line indicates the Fe line profile from the reflection continuum. 
}
\label{fig:Felinefigs}
\end{center}
\end{figure*}

\subsection{GX 3+1}
The \nustar\ data were modeled in the $3.0-20.0$ keV range. The continuum model includes a multi-temperature blackbody, single-temperature blackbody, and power-law component. 
 This provides a poor fit ($\chi^{2}/dof=1491.4/135$) to the \nustar\ data due to the presence of reflection features that are not accounted for by this model.
The Fe line profile from the data can be seen in Figure \ref{fig:Felinefigs}. We utilize the self-consistent thermal reflection model of {\sc relxillns} to properly describe these features. This improves the fit significantly to $\chi^{2}/dof=199.0/128$ ($15\sigma$ improvement via F-test), though this is still a poor statistical fit overall since the high S/N for this source places the data in the systematic and calibration limited regime.  The power-law component is still statistically needed at the 9.5$\sigma$ level of confidence. 
The disk blackbody normalization implies an incredibly small radius ($\sim2.5$ km), even after applying color corrections. This is likely to be a result of spectral hardening of pure blackbody emission by an atmosphere (\citealt{london86}; \citealt{shimura95}; \citealt{merloni00}).  We replace the {\sc diskbb} \& {\sc power-law} components with {\sc nthcomp}, setting the photon seed input to a disk blackbody.  This improves the overall fit further ($\Delta\chi^{2}=34.1$) implying a high optical depth of $\tau\sim7$, but does not change the results for important parameters, i.e., \rin\ and inclination.
The values for reflection model fitting are provided in Table \ref{tab:gx3p1}.  Note that {\sc relxillNS} still contains a single-temperature blackbody continuum component in Model 2. The spectral components and ratio of the data to the overall model can be seen in panel (a) and (b) of Figure \ref{fig:spectra}. 

 There is a flavor of {\sc relxill}, known as {\sc relxillCp}, that allows for reflection from a Comptonized disk component. This model has a hard-coded seed photon temperature of 0.05 keV. This is not appropriate for most NS spectral states, but it was recently employed successfully to model the accreting millisecond pulsar SAX J1808.4-3658 by \citet{disalvo19}. Since we find a low seed photon temperature of $\sim0.09$ keV in our continuum fit with {\sc nthcomp}, we attempt to use {\sc relxillCp} instead of {\sc relxillNS}. The overall model used in {\sc xspec} is {\sc tbabs}*({\sc bbody}+{\sc relxillCp}). This provides a significantly worse fit ($\chi^{2}/dof=344.6/130$ or $>9\sigma$ worse). We show these results in Table \ref{tab:relxillcp} and Figure \ref{fig:relxillcp} of the appendix, but do not report on it further.

\begin{table}
\caption{Reflection Modeling of GX 3+1}
\label{tab:gx3p1} 
\begin{center}
\begin{tabular}{lccc}
\hline
Model & Parameter &  Model 1 & Model 2\\
\hline
{\sc tbabs} &$\mathrm{N}_{\mathrm{H}}$ ($10^{22}$ cm$^{-2}$)&$2.42^{\dagger}$&$2.42^{\dagger}$\\
{\sc diskbb}& kT (keV) & $1.84_{-0.01}^{+0.11}$&...\\
&norm&$15.3_{-2.7}^{+0.4}$&...\\
{\sc powerlaw}& $\Gamma$& $3.5\pm0.1$& ...\\
&norm & $1.04_{-0.02}^{+0.09}$&...\\
{\sc nthcomp}& $\Gamma$ & ... & $1.83_{-0.01}^{+0.04}$ \\
& kT$_{e}$ (keV)& ... & $2.45\pm0.04$\\
& kT$_{bb}$ (keV)& ... &$0.09_{-0.08}^{+0.15}$\\
& norm & ... & $0.86\pm0.02$\\
{\sc relxillNS}&$q$ & $3.2_{-0.6}^{+0.1}$& $2.8\pm0.5$\\
&$i$ ($^{\circ}$)& $28_{-1}^{+3}$&$25\pm1$\\
&$R_{in}$ ($R_{\mathrm{ISCO}}$)&$1.8_{-0.6}^{+0.2}$&$2.2\pm0.2$\\
&$R_{in}$ ($R_{g}$)&$10.8_{-3.6}^{+1.2}$&$13.2\pm1.2$\\
&$kT_{bb}$ (keV) &$2.60_{-0.02}^{+0.04}$ & $1.6\pm0.1$ \\
&$\log \xi$&$2.75_{-0.10}^{+0.05}$&$3.2_{-0.2}^{+0.1}$\\ 
&$A_{Fe}$&$0.62_{-0.03}^{+0.17}$&$3\pm2$\\
&$\log n$ (cm$^{-3}$)&$16.6_{-0.3}^{+0.2}$&$16.1_{-0.1}^{+0.2}$\\
&$f_{\mathrm{refl}}$&$0.93_{-0.01}^{+0.04}$&$0.39\pm0.07$\\
&norm ($10^{-3}$)&$0.72_{-0.12}^{+0.05}$&$1.2\pm0.2$\\
&F$_{\mathrm{unabs}}$&$8_{-2}^{+1}$ &$6.5\pm1.1$\\
&L ($10^{37}$ \lumcgs)&$4.1_{-1.0}^{+0.5}$&$3.3\pm0.6$\\
&L/L$_{\mathrm{Edd}}$& 0.11 & 0.10\\
\hline
&$\chi^{2}$ (dof)& 199.0 (128) & 164.9 (128)\\ 
\hline
{$^{\dagger}=\mathrm{fixed}$}
\end{tabular}

\medskip
Note.---  Errors are reported at the 90\% confidence level and calculated from Markov chain Monte Carlo (MCMC) of chain length 500,000. Data were fit in the $3.0-20.0$ keV band. The outer disk radius was fixed at 990 $R_{g}$, the dimensionless spin parameter and redshift were set to zero for the {\sc relxillNS} model. $f_{\mathrm{refl}}$ denotes the reflection fraction. The unabsorbed flux is taken in the $0.5-50.0$ keV band and given in units of $10^{-9}$ \fluxcgs. Luminosity is calculated based upon a distance of  $D_{\mathrm{max}}=6.5$ kpc \citep{galloway08}.
\end{center}
\end{table} 

\subsection{4U 1702$-$429}
The \nustar\ data were modeled in the $3.0-50.0$ keV energy band. The 
continuum is well described by a cutoff power-law alone ($\chi^{2}/dof=1162.4/388$), but there is a broad Fe line feature present in the ratio of the data to the continuum (Figure \ref{fig:Felinefigs}). In order to accommodate this feature, we apply the standard version of the self-consistent reflection model {\sc relxill}. We initially fix the iron abundance at twice solar since the fit tended towards the maximum value of $A_{Fe}=10$.  Applying the reflection model with $A_{Fe}=2$ provides a $\sim17\sigma$ improvement in the overall fit ($\chi^{2}/dof=526.6/383$)  in comparison the continuum only model.  Last, we explore the dependence on the iron abundance by allowing it to be a free parameter. The value is greater than 4.6 times the solar value, but the position of the inner disk is consistent with the fit that had a fixed $A_{Fe}$.   Both fits are reported in Table \ref{tab:4u1702}. This overabundance could be indicative of a higher density disk than the hard coded value of $10^{15}$ cm$^{-3}$ in {\sc relxill}. We note that the model flavor {\sc relxillD} provides the option of variable density in the disk. However, the current version of this model has a fixed cutoff energy of 300 keV, which is much higher than the value required to fit these data.
Panel (c) of Figure \ref{fig:spectra} shows the model components and ratio of the \nustar\ fit with $A_{Fe}$ left free to vary. Additionally, we also try fitting the spectrum with {\sc relxillCP}, but do not find an improvement in the overall fit ($\Delta\chi^{2}$ increases by 4.7 for the same number of degrees of freedom). We present this in Table \ref{tab:relxillcp} and Figure \ref{fig:relxillcp} in the appendix.

\begin{table}
\caption{Reflection Modeling of 4U 1702-429}
\label{tab:4u1702} 
\begin{center}
\begin{tabular}{lccc}
\hline
Model & Parameter &Fixed $A_{Fe}$ & Free $A_{Fe}$ \\

\hline
{\sc tbabs} &$\mathrm{N}_{\mathrm{H}}$ ($10^{22}$ cm$^{-2}$)&$2.32^{\dagger}$&$2.32^{\dagger}$\\
{\sc relxill}&$q$ & $2.5_{-0.3}^{+1.2}$ & $2.5_{-0.1}^{+1.4}$\\
&$i$ ($^{\circ}$)& $59_{-6}^{+5}$&$61_{-14}^{+2}$\\
&$R_{in}$ ($R_{\mathrm{ISCO}}$)&$1.5_{-0.4}^{+1.6}$&$1.6_{-0.1}^{+2.9}$\\
&$R_{in}$ ($R_{g}$)& $9.0_{-2.4}^{+9.6}$ & $9.6_{-0.6}^{+17.4}$ \\
&$\Gamma$ &$1.97_{-0.04}^{+0.02}$ &$1.97_{-0.03}^{+0.02}$ \\
&$\log \xi$&$3.74_{-0.03}^{+0.25}$&$4.02_{-0.03}^{+0.33}$\\ 
&$A_{Fe}$&$2.0^{\dagger}$& $4.9_{-0.3}^{+4.6}$\\
&E$_{\mathrm{cut}}$&$53_{-2}^{+11}$& $57_{-4}^{+9}$\\
&$f_{\mathrm{refl}}$ &$0.57_{-0.01}^{+0.92}$& $0.5_{-0.2}^{+0.1}$\\
&norm ($10^{-3}$)&$0.70_{-0.30}^{+0.04}$& $0.75_{-0.05}^{+0.09}$ \\
&F$_{\mathrm{unabs}}$&$0.58_{-0.25}^{+0.03}$ &$0.57_{-0.04}^{+0.07}$\\
&L ($10^{36}$ \lumcgs)&$2.2_{-0.9}^{+0.1}$&$2.2_{-0.1}^{+0.3}$\\
&L/L$_{\mathrm{Edd}}$&0.006 & 0.006 \\
\hline
&$\chi^{2}$ (d.o.f.)& 526.6 (383) & 480.2 (382)\\ 
\hline
{$^{\dagger}=\mathrm{fixed}$}
\end{tabular}

\medskip
Note.---  Errors are reported at the 90\% confidence level and calculated from Markov chain Monte Carlo (MCMC) of chain length 500,000. \nustar\ data were fit in the $3.0-50.0$ keV band. The outer disk radius was fixed at 990 $R_{g}$, the dimensionless spin parameter and redshift were set to zero for the {\sc relxill} model. $f_{\mathrm{refl}}$ denotes the reflection fraction. The unabsorbed flux is taken in the $0.5-50.0$ keV band and given in units of $10^{-9}$ \fluxcgs. Luminosity is calculated based upon a distance of $D_{\mathrm{max}}=5.65$ kpc \citep{galloway08}.
\end{center}
\end{table}

\subsection{4U 0614+091}
The \nustar\ data were modeled in the $3.0-30.0$ keV energy band. The column density was fixed at the value inferred from fitting the \xmm/RGS of N$_{\mathrm{H}}=3.46\times10^{21}$ cm$^{-2}$. The continuum is consistent with the hard state, which is well described by a single-temperature blackbody and cutoff power-law ($\chi^{2}/dof=457.3/220$). There is a broad emission feature in the Fe K band that can be seen in Figure \ref{fig:Felinefigs}. We employ {\sc relxill} to account for the reflected emission. This provides a $\sim10\sigma$ improvement in the overall fit. Parameter values are given in Table \ref{tab:4u0614}.  The spectrum and spectral components can be seen in panel (d) of Figure \ref{fig:spectra}.  

\begin{table}
\caption{Reflection Modeling of 4U 0614+091}
\label{tab:4u0614} 
\begin{center}
\begin{tabular}{lcc}
\hline
Model & Parameter & {\it NuSTAR} \\
\hline
{\sc tbabs} &$\mathrm{N}_{\mathrm{H}}$ ($10^{21}$ cm$^{-2}$)&$3.46^{\dagger}$\\
{\sc bbody}& kT (keV) &  $1.51_{-0.01}^{+0.03}$\\
&norm ($10^{-3}$)& $3.55_{-0.42}^{+0.04}$\\
{\sc relxill}&$q$ & $2.07_{-0.04}^{+0.50}$\\
&$i$ ($^{\circ}$) & $52_{-2}^{+10}$\\
&$R_{in}$ ($R_{\mathrm{ISCO}}$)& $1.3_{-0.2}^{+5.4}$ \\
&$R_{in}$ ($R_{g}$)& $7.8_{-1.2}^{+32.4}$ \\
&$\Gamma$ &  $2.57_{-0.24}^{+0.03}$\\
&$\log \xi$& $3.35_{-0.06}^{+0.12}$\\ 
&$A_{Fe}$& $0.58_{-0.06}^{+0.86}$ \\
&E$_{\mathrm{cut}}$ & $16_{-4}^{+1}$\\
&$f_{\mathrm{refl}}$ & $1.0_{-0.4}^{+0.3}$\\
&norm ($10^{-3}$) & $4.1_{-1.1}^{+0.6}$ \\
&F$_{\mathrm{unabs}}$ & $2.2_{-0.6}^{+0.3}$\\
&L ($10^{36}$ \lumcgs)& $2.7_{-0.7}^{+0.4}$ \\
&L/L$_{\mathrm{Edd}}$ & 0.007 \\
\hline
&$\chi^{2}$ (d.o.f.)& 249.8 (213) \\ 
\hline
{$^{\dagger}=\mathrm{fixed}$}
\end{tabular}

\medskip
Note.---  Errors are reported at the 90\% confidence level and calculated from Markov chain Monte Carlo (MCMC) of chain length 500,000. The data were fit in the $3.0-30.0$ keV band. The outer disk radius was fixed at 990 $R_{g}$, the dimensionless spin parameter and redshift were set to zero for the {\sc relxill} model. $f_{\mathrm{refl}}$ denotes the reflection fraction. The unabsorbed flux is taken in the $0.5-50.0$ keV band and given in units of $10^{-9}$ \fluxcgs.  Luminosity is calculated based upon a distance of $D_{\mathrm{max}}=3.2$ kpc \citep{kuulkers10}.
\end{center}
\end{table} 

\subsection{4U 1746$-$371}
The \nustar\ data were modeled in the $3.0-25.0$ keV band. The column density along the line of sight is fixed at N$_{\mathrm{H}}=3.76\times10^{21}$ cm$^{-2}$. The spectrum can be entirely described by a cutoff power-law and single temperature blackbody component ($\chi^{2}/dof=224.3/177$). There is not a strong detection of an Fe line component (see Figure \ref{fig:Felinefigs}).  Applying a Gaussian component at 6.4 keV provides a marginal improvement of $\chi^{2}/dof=211.3/175$ (or 2.8$\sigma$) with an equivalent width of $\sim30$ eV.  Allowing the spectrum to be described by reflection with {\sc relxill} provides an improvement in the fit ($\Delta \chi^{2}$ decreases by 17.7 for 2 dof). This is a 3.4$\sigma$ improvement in comparison to the continuum only modeling. We fixed the following parameter values in the reflection model to those typical of other sources: $q=3.2$ \citep{wilkins18}, $R_{in}=1\ R_{\mathrm{ISCO}}$, $A_{Fe}=1$, $\log(\xi)=3.0$, and an inclination of $i=75^{\circ}$ (because it is a \lq \lq dipping" source). We can then place an upper limit on the presence of reflection to be 13\% from the reflection fraction. Allowing the fixed parameters within {\sc relxill} to be free does not provide any meaningful constraints. For example, the inner disk radius is completely unconstrained (i.e., consistent with both hard-coded limits of 1~\risco\ and 100~\risco). This was also the case for the ionization parameter which was consistent with $\log{\xi}=0$ and $\log{\xi}=4.7$. We therefore are unable to learn more information about the reflected component.

\begin{figure*}
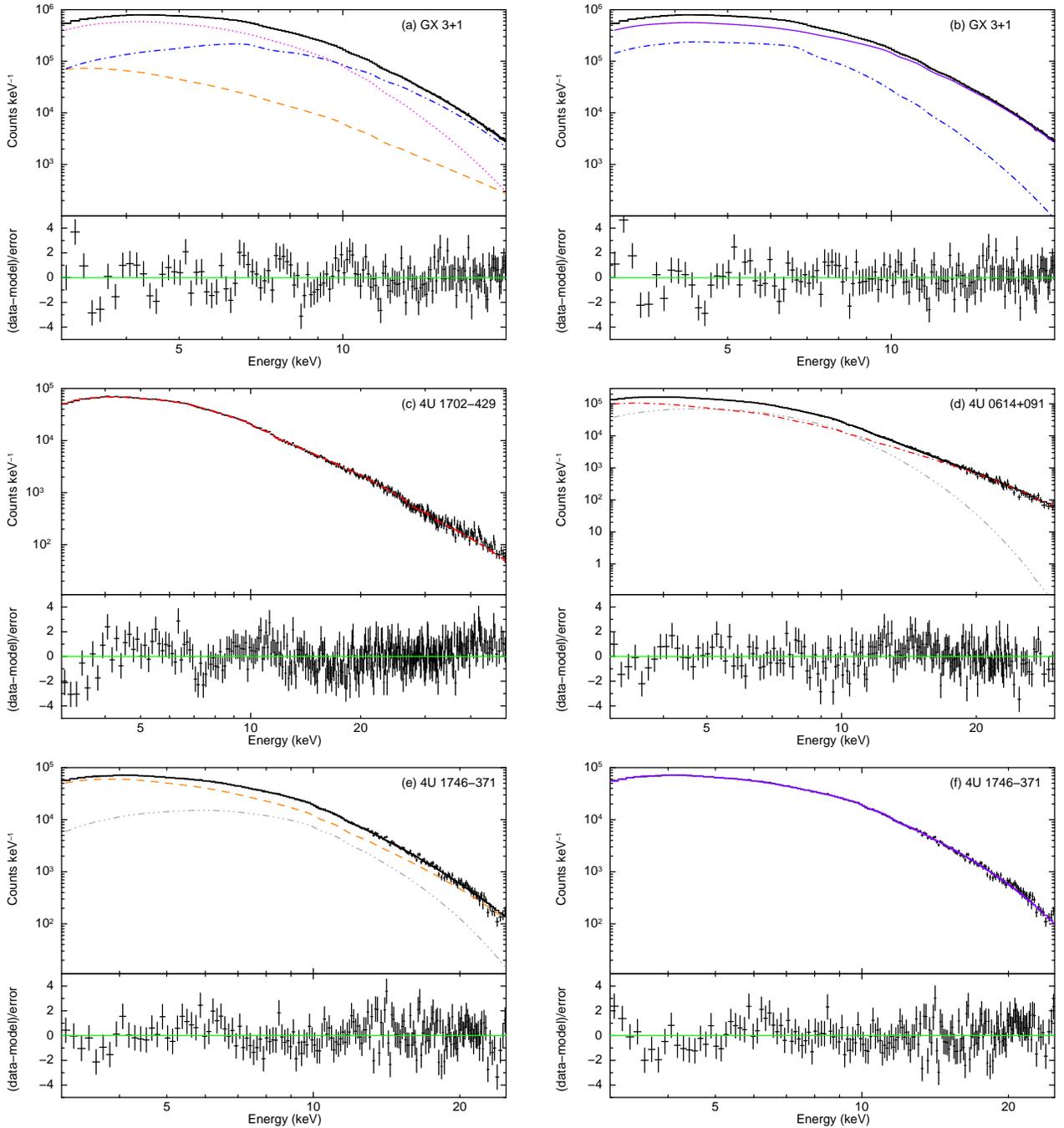
 
\begin{center}
\includegraphics[width=6.cm, angle=270]{gx3p1_plbbrel_11072018.eps}
\includegraphics[width=6.cm, angle=270]{gx3p1_nthrel_11072018.eps}
\includegraphics[width=6.cm, angle=270]{4u1702_freeAFe_11072018.eps}
\includegraphics[width=6.cm, angle=270]{4u0614_bbrel_11072018.eps}
\includegraphics[width=6.cm, angle=270]{4u1746_bbpow_11072018.eps}
\includegraphics[width=6.cm, angle=270]{4u1746_nth_11072018.eps}
\caption{\nustar\ spectral modeling and residuals divided by errors for the sample of atoll sources. Dot-dashed lines indicate the reflection model (blue: {\sc relxillns}, red: {\sc relxill}). The magenta dotted line is a multi-temperature blackbody to account for disk emission. The grey dot-dot-dot dashed line is a single-temperature blackbody to account for boundary layer emission. The orange dashed line indicates the power-law component. The purple solid line is the Comptonization model. See Tables 2-5 for component parameters. }
\label{fig:spectra}
\end{center}
\end{figure*}

The source is likely highly Comptonized. Switching out the continuum for Comptonization alone ({\sc nthcomp}: \citealt{nthcomp1}; \citealt{nthcomp2}) provides a comparatively good fit  ($\chi^{2}/dof=226.4/178$) and implies a high optical depth of $\tau\sim6$. Both continuum model fits are reported in Table \ref{tab:4u1746}.  The spectrum and spectral components for each fit can be seen in panels (e) and (f) of Figure \ref{fig:spectra}.  We try to apply {\sc relxillCp} to again check the presence of reflection in this observation. We fix the same parameters as the fit performed with {\sc relxill} while allowing the continuum parameters, reflection fraction, and normalization to vary. This provides an improvement over the spectral modeling with {\sc nthcomp} alone ($\Delta\chi^{2}$=11 for the same number of dof). We obtain a higher upper limit on the presence of reflection ($\sim56$\%), but we are still unable to constrain reflection parameter values when they are allowed to vary. The combination of high inclination and strong Comptonization could be responsible for scattering any potential reflection features in this system out of the line of sight (see Figure 5 of \citealt{petrucci01}) and suggests we are observing through the Comptonizing corona that is on top of the accretion disk.

\begin{table}
\caption{Continuum Modeling of 4U 1746-371}
\label{tab:4u1746} 
\begin{center}
\begin{tabular}{lccc}
\hline
Model & Parameter & Power-law & Comptonization\\
\hline
{\sc tbabs} &$\mathrm{N}_{\mathrm{H}}$ ($10^{21}$ cm$^{-2}$)&$3.76^{\dagger}$&$3.76^{\dagger}$\\
{\sc bbody}& kT (keV) & $2.4\pm0.2$ & ...\\ 
& norm ($10^{-4}$) & $5.4\pm0.9$ & ...\\
{\sc cutoffpl}& $\Gamma$ & $1.3\pm0.2$ & ...\\
& E$_{\mathrm{cut}}$ & $5.5\pm0.7$ & ...\\
& norm ($10^{-2}$)& $6.6\pm0.7$ & ...\\
{\sc nthcomp}& $\Gamma$&...&$1.91\pm0.02$\\
& kT$_{\mathrm{e}}$ (keV) &...&$2.91\pm0.04$\\
& kT$_{\mathrm{bb}}$ (keV)&...&$0.50_{-0.05}^{+0.04}$\\
&norm ($10^{-2}$)&...&$2.4_{-0.3}^{+0.4}$\\
&F$_{\mathrm{unabs}}$& $0.45\pm0.09$ &$0.34_{-0.04}^{+0.06}$\\
&L ($10^{36}$ \lumcgs)& $8\pm2$ &$5.8_{-0.7}^{+1.0}$\\
&$\mathrm{L}/\mathrm{L}_{\mathrm{Edd}}$ & $0.02$ & $0.015$ \\ 
\hline
&$\chi^{2}$ (d.o.f.)& 224.3 (177) & 226.4 (178)\\ 
\hline
{$^{\dagger}=\mathrm{fixed}$}
\end{tabular}

\medskip
Note.---  Errors are reported at the 90\% confidence level and calculated from Markov chain Monte Carlo (MCMC) of chain length 500,000. \nustar\ was fit in the $3.0-25.0$ keV band. The unabsorbed flux is taken in the $0.5-50.0$ keV band and given in units of $10^{-9}$ \fluxcgs.  Luminosity is calculated based upon a distance of  $D_{\mathrm{max}}=11.9$ kpc \citep{pritzl01}.
\end{center}
\end{table}

\section{Discussion}
We perform a time-averaged spectral analysis with a focus on detecting reflection features in a sample of accreting atoll sources that were granted $\sim20$ ks observations per source with \nustar\ to determine the position of the inner disk. The sources span a range in Eddington fraction from $0.006-0.11$ (assuming the empirical limit of $3.8\times10^{38}$ \lumcgs).  Broad Fe lines were detected in three out of four of the sources: \mbox{GX 3+1}, \mbox{4U~1702$-$429}, and \mbox{4U~0614+091}.  We account for reflection by using different flavors of {\sc relxill} based on the illuminating continuum in these systems. There are other families of reflection models, such as {\sc reflionx} \citep{ross05} or {\sc bbrefl} \citep{ballantyne04}, but \citet{ludlam17a} demonstrated that reflection fitting with these models provide similar results as {\sc relxill}. The final source, \mbox{4U 1746$-$371}, does not require reflection in order to describe the spectrum. However, we can place an upper limit on its presence through the reflection fraction to be $13\%-56\%$ depending on the choice of the illuminating continuum. The ionization parameters of the three sources with reflection are consistent with $\log \xi=2.3-4.0$ seen in other NS LMXBs (\citealt{cackett10}; \citealt{ludlam17a}). 

\mbox{GX 3+1} displayed a particularly soft spectrum that was well described by a double thermal model with a power-law component. The reflection model determined the accretion disk was truncated prior to the ISCO at $1.8_{-0.6}^{+0.2}\ R_{\mathrm{ISCO}}$ ($10.8_{-3.6}^{+1.2}\ R_{g}$) and inclination of $27^{\circ}-31^{\circ}$. The inner disk position agrees with the $\sim10\ R_{g}$ limit found by \citet{pintore15}, but the inclination is slightly lower than previous investigations with {\it XMM-Newton} by \citet{piraino12} and \citet{pintore15}. A potential source of the discrepancy between the measured inclinations could be due to the difference in N$_{\mathrm{H}}$ used when performing spectral fits.  The choice of absorption model used (e.g., {\sc wabs}, {\sc phabs}, {\sc tbabs}, etc.), as well as abundances and cross-sections adopted can result in changes in the measured N$_{\mathrm{H}}$ by up to 30\% (Gatuzz \& Garc\'{i}a, in prep.). This can alter the edge at 7 keV in the Fe line region, which can impact the blue wing predicted by the reflection model when fitting, and therefore lead to changes in the inferred inclination. Additionally, we obtain similar values for \rin\ and inclination when using a Comptonized disk blackbody component for the continuum.

The position of the inner disk was close to the ISCO for \mbox{4U~1702$-$429}, though the large error bars are also compatible with disk truncation.  This is in agreement with the values reported in \citet{mazzola18} from archival \xmm, \integral, and \beppo\ observations. The iron abundance had to be fixed during the reflection fits. Allowing the iron abundance vary required $>5\times$ the solar abundance. This improved the fit, but the disk was highly ionized ($\log \xi\simeq4$). The super-solar abundance likely indicates that the disk has a higher density than accounted for by the {\sc relxill} model ($10^{15}$ cm$^{-3}$). This behavior was also seen for the BH HMXB Cyg X-1 \citep{tomsick18} which had a super-solar Fe abundance when modeled with a disk density of $10^{15}$ cm$^{-3}$. When the disk density was allowed to increase, the Fe abundance in the reflection model decreased. See \citet{garcia18} for a recent discussion on the relationship between disk density and inferred Fe abundance. 
The inclination is between $53^{\circ}-64^{\circ}$  depending on the value of $A_{Fe}$, in agreement with the results from \citet{iaria16} from \xmm\ and {\it INTEGRAL} data. Additionally, the upper limits on the inner disk position when $A_{Fe}=2.0$ are consistent with the values reported in \citet{iaria16}.

4U~0614$-$091 was also truncated slightly outside of the ISCO. The sub-solar Fe abundance agrees with previous studies and the nature of the donor in this system (\citealt{madej10}; \citeyear{madej14}), although the abundance is consistent with solar at the 90\% confidence level. The inclination of the system from {\sc relxill} is $50^{\circ}-62^{\circ}$, which again agrees with the results from the \xmm\ RGS and EPIC-pn studies performed by Madej et al.\ (2010, 2014). This supports the idea that relativistically blurred Fe and O lines can originate from very similar regions in the disk providing additional diagnostics for the inner accretion flow in these systems \citep{ludlam16}. Future studies conducted with observatories such as \nicer\ \citep{gendreau12} that are sensitive to Fe lines and lower energy emission features simultaneously can confirm if these locations are the same or mutually exclusive.  

Further observations of \mbox{4U~1702$-$429} and \mbox{4U~0614$-$091} would yield tighter constraints on the position of the inner disk radius to determine if the accretion flow is truncated or close to the NS, particularly if \mbox{4U~1702$-$429} is targeted in a state with higher intensity. Additionally, the development of a cutoff power-law reflection model with variable disk density and cutoff energy will shed light on the peculiar super-solar iron abundances implied in the reflection fitting of \mbox{4U~1702$-$429}. The current version of {\sc relxillD} that allows for variable disk density has a fixed cutoff energy of 300 keV.

4U~1746$-$371 did not show a clear signature of reflection in its spectrum. The low cutoff energy is consistent with cutoff energies seen in other NS LMXBs in intermediate states (4U 1636$-$536, \mbox{4U 1705$-$44}, 4U 1728$-$34, 4U 1734$-$44, and 4U 1820$-$30: \citealt{church14}) and an {\it INTEGRAL} study of this source by \citet{balman09}. Although, we find a hotter blackbody component and harder spectral index in comparison to \citet{balman09}. The spectral index is closer to the value of $\Gamma=1.20\pm0.27$ found in \citet{church01}. 
Additionally, we model the continuum emission with a Comptonized accretion disk component. The optically thick Comptonized component ($\tau\sim6$) and low electron temperature ($\sim3$ keV) of 4U~1746$-$371 is akin to GX 13+1 (\citealt{iaria14}; \citealt{dai14}), which is another highly-inclined ``soft" atoll. In contrast to 4U~1746$-$371, GX 13+1 continuously shows a broad Fe line component and narrow Fe absorption features indicative of winds \citep{diaztrigo12}. This could be due to the fact that Fe line flux is correlated with continuum flux \citep{lin10} and GX 13+1 is more than six times as luminous \citep{dai14}. A simple Gaussian at 6.4 keV to account for a potential Fe line component in 4U~1746$-$371 gave an equivalent width of $\sim30$ eV, which is about $1/6^{\mathrm{th}}$ of the value reported in \citet{diaztrigo06}.  However, the observation reported in \citet{diaztrigo06} occurred when the source was a factor of four times more luminous.  The discrepancy in the equivalent width of the Fe line component is likely the result of a change in ionization state, though we were unable to place any meaningful constraints on the ionization parameter with {\sc relxill} or {\sc relxillCp}.

Another source was recently found to not have reflection features present in its persistent spectrum: the Z source \mbox{GX 5$-$1} \citep{homan18}. The likeliest explanation for the absence of reflection features was a highly ionized disk given the sources high luminosity (L$_{1-100\ \mathrm{keV}}=2.97\times10^{38}-4.27\times10^{38}$ \lumcgs). A highly ionized accretion disk could explain the lack of reflection in \mbox{4U 1746$-$371}, but the source is not nearly as luminous as GX 5$-$1. \mbox{4U 1746$-$371} was at 0.02 L$_{\mathrm{Edd}}$ at the time of the \nustar\ observation, whereas GX 5$-$1 spanned the range of $\sim0.76-1.14$ L$_{\mathrm{Edd}}$.
Since \mbox{4U 1746$-$371} is a known dipping source, it is more likely that the combination of source geometry and Comptonization in the accretion disk corona scatters reflected photons out of the line of sight vastly reducing the number of line contributing photons \citep{petrucci01}. This is further supported by the high optical depth ($\tau\sim6$) from the Comptonization fit in addition to the dipping nature of the source. Note that this is not the case for GX~3+1 since the source is at a lower inclination and we are not looking through the Comptonizing corona on top of the accretion disk. However, this could also explain the lack of a reflection spectrum in the \lq \lq soft" state for \mbox{GS~1826$-$24}, which presents evidence of being highly inclined and Comptonized as well \citep{chenevez16}. 

\begin{figure} 
\centering
\includegraphics[width=0.48\textwidth]{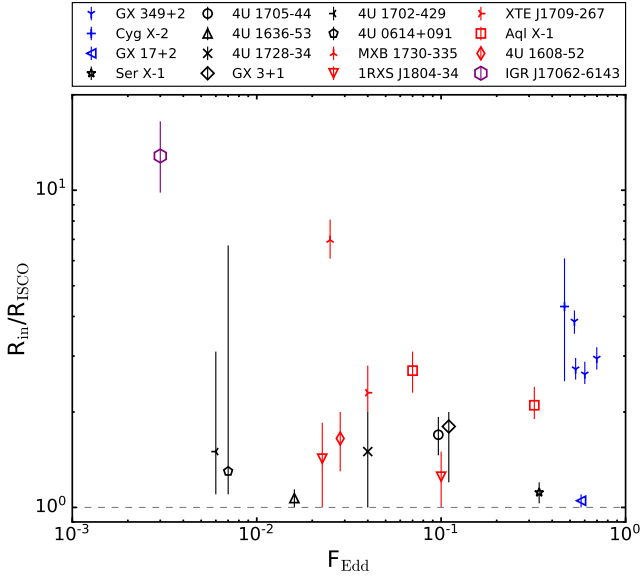}
\caption{Comparison of Eddington fraction and measured inner disk radii for NSs observed with $\emph{NuSTAR}$. Black points indicate persistent atoll sources, blue points indicate persistent Z sources, red points are transient systems, and the purple point indicates a very faint X-ray binary (VFXB). The dashed grey line denoted the innermost stable circular orbit. See Table \ref{tab:edd} for inner disk radii and Eddington fractions. 
}
\label{fig:fedd}
\end{figure}

\begin{table}
\caption{NS Inner Disk Radii \& Eddington Fraction Observed with $\emph{NuSTAR}$ }
\label{tab:edd} 
\begin{center}
\begin{tabular}{lccc}
\hline
Source & $R_{in}$ (ISCO) & F$_{\mathrm{Edd}}$ & ref.\\
\hline
\multicolumn{4}{l}{ Atolls }\\
\hline
Ser X-1 & $1.03-1.20$&0.34  & 1\\
& $1.76-2.70$ & 0.34 & 2\\
4U 1705$-$44 & $1.46-1.93$ & 0.10 & 3\\
4U 1636$-$53 & $1.00-1.14$ & 0.01 & 3\\
4U 1728$-$34 & $1.0-2.0$ & 0.04 & 4\\
GX 3+1& $1.2-2.0$ & 0.11& ...\\
4U 1702$-$429 & $1.1-3.1$ & 0.006&...\\
4U 0614+091& $1.1-5.7$ & 0.007& ...\\
\hline
\multicolumn{4}{l}{ Z }\\
\hline
GX 17+2 & $1.00-1.30$ & 0.57 & 3 \\
Cyg X-2 & $2.5-6.1$ & 0.47 & 5\\
GX 349+2& $2.71-2.96$ & 0.54 & 6\\
& $3.53-4.18$& 0.53 & 6\\
& $2.45-2.88$ & 0.60 & 6\\
& $2.72-3.2$ & 0.70 & 6\\
\hline
\multicolumn{4}{l}{ Transients }\\
\hline
4U 1608$-$52 & $1.3-2.0$ & 0.03 & 7\\
1RXS J180408.9-34 & $1.00-1.85$ & 0.02 & 8\\
& $1.0-1.5$ & 0.10 & 9\\
Aquila X-1 & $2.3-3.1$ & 0.07 & 10\\
& $1.9-2.4$ & 0.23 & 10\\
XTE J1709$-$267& $2.0-2.8$ & 0.04& 11 \\
MXB 1730$-$335& $6.08-8.08$& 0.025 & 12\\
\hline
\multicolumn{4}{l}{ VFXB }\\
\hline
IGR J17062$-$6143& $9.8-16.5$ & $\sim0.003$ & 13\\
\hline
\end{tabular}

\medskip
Note.--- (1) \citealt{miller13};  (2) \citealt{matranga17}; (3) \citealt{ludlam17a}; (4) \citealt{sleator16}; (5) \citealt{mondal18};  (6) \citealt{coughenour17}; (7) \citealt{degenaar15}; (8) \citealt{ludlam16}; (9) \citealt{degenaar16}; (10) \citealt{ludlam17c};   (11) \citealt{ludlam16}; (12) \citealt{vandeneijnden17}; (13) \citealt{vandeneijnden18}.
\end{center}
\end{table}

For the three sources in which reflection was detected, the brightest of the sources is truncated prior to the ISCO whereas the lower Eddington luminosities have inner disk radii nearly consistent with the ISCO, within errors. We are able to place these three sources into the larger sample sof NSs (with $B<10^{10}$ G) that have been observed with \nustar\ wherein reflection allows limits to be placed on the location of the disk. Table \ref{tab:edd} and Figure \ref{fig:fedd} are updated from \citet{ludlam17a} to incorporate recent analyses and further division into NS sub-classes (persistent atoll and Z sources, transients, and a very faint X-ray binary). The Eddington fraction, which is a proxy for \mdot, is calculated from the 0.5-50 keV luminosity of each source divided by the empirical Eddington limit of $3.8\times10^{38}$ \lumcgs. The observations of \mbox{4U~1702$-$429} and \mbox{4U~0614+091} in this work provide the lowest luminosity disk position measurements for persistent atoll sources observed with \nustar. In the case of the very faint X-ray binary, \mbox{IGR~J17062$-$6143} is able to constantly accrete at very low \mdot\ where the disk may enter the radiatively inefficient accretion flow (RIAF) regime (\citealt{narayan94}; \citealt{blandford99}), but truncation by the magnetosphere is not ruled out \citep{vandeneijnden18}. 

It should be noted that the \rin\ estimates for Serpens~X-1 by \citet{miller13} and \citet{matranga17} used the same \nustar\ data set, however \citet{matranga17} also used \xmm\ observations to construct a broad-band X-ray spectrum. This emphasizes the importance of the lower-energy bandpass in deriving reflection parameters. Hence values in Table 6 and Figure 3 may be biased since \nustar\ only has the high energy coverage. Regardless, the lack of a clear trend between F$_{\mathrm{Edd}}$ and $R_{in}$ reaffirms the  previously report complex behavior of the disk over various mass accretion rates (\citealt{cackett10}; \citealt{Dai10}; \citealt{ludlam17a}) and suggests the presence of a boundary layer or the magnetic field of the NS likely plays a role. 

\begin{table}
\caption{Maximum Boundary Layer Extent and Magnetic Field Strength}
\label{tab:truncation} 
\begin{center}
\begin{tabular}{lccc}
\hline
Source & $R_{\mathrm{BL,\ max}}$ ($R_{g}$) & $B_{\mathrm{max}}$ ($10^{8}$ G) \\
\hline
GX 3+1& $\sim6.67$ & $\leq6.7$\\
4U 1702$-$429& $\sim5.35$ & $\leq3.3$ \\
4U 0614+091& $\sim5.36$ & $\leq14.5$\\
\hline
\end{tabular}

\medskip
Note.--- The maximum radial extent of the boundary layer region is calculated based up the maximum luminosity reported in Table $2-4$. We assume a canonical NS ($M_{NS}=1.4\ M_{\odot}$, $R_{NS}=10$ km). For the estimate of the upper limit on the magnetic field strength at the poles, we use the maximum unabsorbed $0.5-50$ keV flux and inner disk radius from the \nustar\ fits. Additionally, we assume an angular anisotropy and conversion factor of unity \citep{cackett09}. The efficiency of accretion in both calculations is assumed to be 0.2 \citep{ss00}. 
\end{center}
\end{table}

In Table \ref{tab:truncation} we provide estimates for the extent of a boundary layer using Equation (25) from \citet{ps01} and upper limits on the magnetic field strength using Equation (1) from \citet{cackett09} for these systems. 
The large upper limit on the magnetic field strength for 4U~0614+091 is driven by the large uncertainty on the inner disk radius. Although we do not detect pulsations during these observations, the sources could still be magnetically accreting. 
The hot spot could be nearly aligned with the spin axis, in which case pulsations would be undetectable, or the modulated emission could be scattered by the circumstellar gas \citep{lamb85}.  A comprehensive explanation regarding all the ways pulsations are suppressed or hidden from view in NS LMXBs can be found in \citet{lamb09}. 
The extent of the boundary layer regions are too small to account for the disk position, but these values may be underestimated as they do not account for spin and viscous effects in this layer.  An additional plausible explanation for disk truncation could be that the innermost region of the accretion disk has given way to a compact Comptonizing coronal region as expected in lower luminosity regimes (\citealt{narayan94}; \citealt{done07}; \citealt{veledina13}). We are unable to determine the exact truncation mechanism from a single observation. This is because the position of the disk as a function of \mdot\ changes in opposing manners for truncation by the magnetosphere \citep{ip09} or a boundary layer extending from the NS surface \citep{ps01}. Multiple observations over significant changes in mass accretion rate within individual sources are needed to determine the definitive truncation mechanism for a system.

\section{Summary}
We present a spectral analysis of four persistent atoll sources observed with \nustar\ to investigate the location of the inner disk measured via reflection fitting techniques within these relatively low luminosity systems. We detect the presence of reflection firmly in three out of four of these sources (\mbox{GX 3+1}, \mbox{4U~1702$-$429}, and \mbox{4U~0614+091}).  Reflection features were not detected in \mbox{4U 1746$-$371} likely due to a combination of source geometry and strong Comptonization. These sources span a range in Eddington fraction of $0.006-0.11$, providing the lowest F$_{\mathrm{Edd}}$ disk position measurement for an atoll source observed with \nustar, and increase the number of sources with detected reflection features by $\sim20$\% for the \nustar\ sample. Adding these sources to the existing sample of NSs with inner radius measurements reaffirms the lack of a clear one-to-one trend between the position of the inner disk and mass accretion rate as observed with other X-ray missions such as \xmm\ and {\it Suzaku}  (e.g., \citealt{cackett10}; \citealt{Dai10}; \citealt{chiang16b}). This emphasizes the need to shift focus to investigate the truncation mechanisms in individual sources to determine the dynamical role of the boundary layer or magnetosphere. In order to disentangle these different scenarios of disk truncation, multiple observations of a source over a large range in mass accretion rate are necessary. 

\acknowledgements{We thank the referee for their comments which have improved this work. This research has made use of the NuSTAR Data Analysis Software (NuSTARDAS) jointly developed by the ASI Science Data Center (ASDC, Italy) and the California Institute of Technology (Caltech, USA). R.M.L.\ is funded through a NASA Earth and Space Science Fellowship. E.M.C.\ gratefully acknowledges NSF CAREER award AST-1351222. We thank J. van den Eijnden for verifying the unabsorbed flux of \mbox{IGR~J17062$-$6143}.}

\newpage

\appendix
\subsection*{\small{Spectral fitting with relxillcp}}

\renewcommand{\thefigure}{A\arabic{figure}}
\renewcommand{\thetable}{A\arabic{table}}
Here we provide the reader with the results from using the model {\sc relxillCp} on GX~3+1 and 4U~1702$-$429. This model produces reflection from the reprocessing of photons from a Comptonized disk component with a hard-coded seed photon temperature of 0.05 keV. Table \ref{tab:relxillcp} provides the parameter values that can be compared to the resulting fits in Tables 2 and 3 for GX~3+1 and 4U~1702$-$429, respectively. Figure \ref{fig:relxillcp} is given to provide a direct comparison to the fits presented in Figure \ref{fig:spectra}. These fits do not provide an improvement in the overall fit of the spectra from the results presented in \S 3.1 and \S 3.2. In the case of 4U~1702$-$429, the comparable fit with high $A_{Fe}$ and ionization still supports the need for higher disk density models. 

\begin{table*}[h!]
\caption{Reflection Modeling with {\sc relxillCp}}
\label{tab:relxillcp} 
\begin{center}
\begin{tabular}{lccc}
\hline
Model & Parameter & GX 3+1 & 4U 1702$-$429\\
\hline
{\sc tbabs} &$\mathrm{N}_{\mathrm{H}}$ ($10^{22}$ cm$^{-2}$)&$2.42^{\dagger}$&$2.32^{\dagger}$\\
{\sc bbody}& kT (keV) & $1.40\pm0.02$&...\\
&norm  ($10^{-3}$)&$6.8\pm0.3$&... \\
{\sc relxillCP}&$q$ & $3.1\pm0.3$& $2.3\pm0.2$\\
&$i$ ($^{\circ}$)& $24\pm1$&$64\pm1$\\
&$R_{in}$ ($R_{\mathrm{ISCO}}$)&$1.85_{-0.27}^{+0.35}$&$1.00^{+0.4}_{*}$\\
&$R_{in}$ ($R_{g}$)&$10.8_{-1.6}^{+2.1}$&$6.0^{+2.4}_{*}$\\
&$\Gamma$ & $1.72_{-0.01}^{+0.03}$ & $2.00\pm0.01$\\
&$\log \xi$&$3.30_{-0.05}^{+0.01}$&$4.7_{-0.1}^{*}$\\ 
&$A_{Fe}$&$10_{-0.2}^{*}$&$8.7_{-0.9}^{+0.7}$\\
&$kT$ (keV) &$2.24\pm0.01$ & $20_{-1}^{+2}$ \\
&$f_{\mathrm{refl}}$&$0.66_{-0.07}^{+0.03}$&$2.6_{-0.5}^{+0.4}$\\
&norm ($10^{-3}$)&$7.2\pm0.1$&$0.30_{-0.07}^{+0.02}$\\
\hline
&$\chi^{2}$ (dof)& 344.6 (130) & 484.9 (382) \\ 
\hline
{$^{\dagger}=\mathrm{fixed}$}
\end{tabular}

\medskip
Note.---  Errors are reported at the 90\% confidence level. An asterisk indicates that the parameter is at the hard-coded limit of the model. The outer disk radius was fixed at 990 $R_{g}$, the dimensionless spin parameter and redshift were set to zero for the {\sc relxillCp} model. $f_{\mathrm{refl}}$ denotes the reflection fraction. 
\end{center}
\end{table*}

\begin{figure*}[h!]
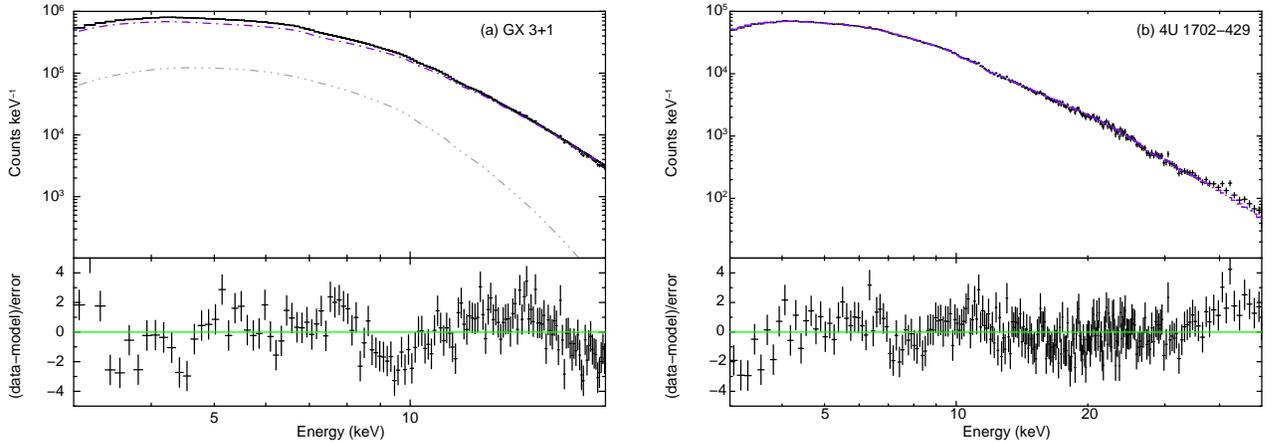

\begin{center}
\includegraphics[width=6.cm, angle=270]{gx3p1relxillcp.eps}
\includegraphics[width=6.cm, angle=270]{4u1702m429relxillcp.eps}
\caption{\nustar\ spectral modeling and residuals divided by errors for (a) GX~3+1 and (b) 4U~1702$-$429 when fit with {\sc relxillCp}. The purple dot-dashed lines indicates {\sc relxillCp}, which contains both the Comptonized disk component of the continuum and reflection. The grey dot-dot-dot dashed line is a single-temperature blackbody to account for boundary layer emission. }
\label{fig:relxillcp}
\end{center}
\end{figure*}


\begin{thebibliography}{}
\bibitem[Arnaud(1996)]{arnaud96}Arnaud, K. A.\ 1996, in Astronomical Society of the Pacific Conference Series, Vol. 101, Astronomical Data Analysis Software and Systems V, ed.\ G. H. Jacoby \& J. Barnes, 17
\bibitem[Asai et al.(2000)]{asai00}Asai, K., Dotani, T., Nagase, F., \& Mitsuda, K.\ 2000, ApJS, 131, 571
\bibitem[Ballantyne(2004)]{ballantyne04} Ballantyne, D. R.\ 2004, MNRAS, 351, 57
\bibitem[Balman(2009)]{balman09}Balman, \c{S}.\ 2009, ApJ, 138, 50
\bibitem[Ba\l uci\'{n}ska-Church et al.(2004)]{bcs04}Ba\l uci\'{n}ska-Church, M., Church, M. J., \& Smale, A. P.\ 2004, MNRAS, 347, 334
\bibitem[Bardeen et al.(1972)]{bardeen72}Bardeen, J. M., Press, W. H., \& Teukolsky, S. A.\ 1972, ApJ, 178, 347
\bibitem[Blandford \& Begelman(1999)]{blandford99}Blandford, R. D., \& Begelman, M. C.\ 1999, MNRAS, 303, L1
\bibitem[Braje et al.(2000)]{braje00}Braje, T. M., Romani, R. W., \& Rauch, K. P.\ 2000, ApJ, 531, 447
\bibitem[Brandt et al.(1992)]{brandt92}Brandt S., Castro-Tirado A. J., Lund N., et al.\ 1992, A\&A, 262, L15
\bibitem[Cackett et al.(2008)]{cackett08}Cackett, E. M., Miller, J. M., Bhattacharyya, S., et al.\ 2008, ApJ, 674, 415
\bibitem[Cackett et al.(2009)]{cackett09}Cackett, E. M., Altamirano, D., Patruno, A., et al.\ 2009, ApJL, 694, L21
\bibitem[Cackett et al.(2010)]{cackett10}Cackett, E. M., Miller, J. M., Ballantyne, D. R., et al.\ 2010, ApJ, 720, 205
\bibitem[Chenevez et al.(2006)]{chenevez06}Chenevez, J., Falanga, M., Brandt, S., et al.\ 2006, A\&A, 449, L5
\bibitem[Chenevez et al.(2016)]{chenevez16}Chenevez, J., Galloway, D. K., in't Zand, J. J. M., et al.\ 2016, ApJ, 818, 135
\bibitem[Chiang et al.(2016a)]{chiang16}Chiang, C.-Y., Cackett, E. M., Miller, J. M., et al.\ 2016a, ApJ, 821, 105
\bibitem[Chiang et al.(2016b)]{chiang16b}Chiang, C.-Y., Morgan, R. A., Cackett, E. M., et al.\ 2016b, ApJ, 831, 45
\bibitem[Choudhury et al.(2017)]{choudhury17}Choudhury, K., Garc\'{i}a, J. A., Steiner, J. F., \& Bambi, C.\ 2017, ApJ, 851, 57
\bibitem[Church et al.(2001)]{church01}Church, M. J., \& Ba\l uci\'{n}ska-Church, M.\ 2001, A\&A, 369, 915
\bibitem[Church et al.(2014)]{church14}Church, M. J., Gibiec, A., \& Ba\l uci\'{n}ska-Church, M.\ 2014, MNRAS, 438, 2784
\bibitem[Cook, Shapiro, \& Teukolsky(1994)]{cook94}Cook, G. B., Shapiro, S. L., \& Teukolsky, S. A.\ 1994, ApJ, 424, 823
\bibitem[Coughenour et al.(2018)]{coughenour17}Coughenour, B. M., Cackett, E. M., Miller, J. M., \& Ludlam, R. M.\ 2018, ApJ, 867, 64
\bibitem[D'A\`{i} et al.(2010)]{Dai10}D'A\`{i}, A., Di Salvo, T., Ballantyne, D., et al.\ 2010, A\&A, 516 A36
\bibitem[D'A\`{i} et al.(2014)]{dai14} D'A\`{i}, A., Iaria, R., Di Salvo, T., Riggio, A., Burderi, L., \& Robba, N. R.\ 2014, A\&A, 564, A62
\bibitem[Degenaar et al.(2015)]{degenaar15}Degenaar, N., Miller, J. M., Chakrabarty, D., et al.\ 2015, MNRAS, 451, L85
\bibitem[Degenaar et al.(2016)]{degenaar16}Degenaar, N., Altamirano, D., Parker, M., et al.\ 2016, MNRAS, 461, 4049
\bibitem[Deutsch et al.(1998)]{deutsch98}Deutsch, E. W., Anderson, S. F., Margon, B., \& Downes, R.\ 1998, ApJ, 493, 775
\bibitem[Di Salvo et al.(2019)]{disalvo19} Di Salvo, T., Sanna, A., Burderi, L., et al.\ 2019, MNRAS, 483, 767
\bibitem[D\'{i}az Trigo et al.(2006)]{diaztrigo06}D\'{i}az Trigo, M., Parmar, A. N., Boirin, L., Mendez, M., \& Kaastra, J.\ 2006, A\&A, 445, 179
\bibitem[D\'{i}az Trigo et al.(2012)]{diaztrigo12} D\'{i}az Trigo, M., Sidoli, L., Boirin, L., M., \& Parmar, A. N.\ 2012, A\&A, 543, A50
\bibitem[Done et al.(2007)]{done07}Done, C., Gierli\'{n}ski, M., \& Kubota, A.\ 2007, A\&A Rev., 15, 1
\bibitem[Evans et al.(2007)]{evans07}Evans, P. A., Beardmore, A. P., Page, K. L., et al.\ 2007, A\&A, 469, 379
\bibitem[Evans et al.(2009)]{evans09}Evans, P. A., Beardmore, A. P., Page, K. L., et al.\ 2009, MNRAS, 397, 1177
\bibitem[Fabian et al.(1989)]{fabian89}Fabian, A. C., Rees, M. J., Stella, L., \& White, N. E.\ 1989, MNRAS, 238, 729
\bibitem[Galloway et al.(2008)]{galloway08}Galloway, D. K., Muno, M. P., Hartman, J. M., Psaltis, D., \& Chakrabarty, D.\ 2008, ApJS, 179, 360
\bibitem[Garc\'{i}a et al.(2014)]{garcia14} Garc\'{i}a, J. A., Dauser, T., Lohfink, A., et al.\ 2014, ApJ, 782, 76
\bibitem[Garc\'{i}a et al.(2018)]{garcia18} Garc\'{i}a, J. A., Kallman, T. R., Bautista, M., et al.\ 2018, arxiv:1805.00581
\bibitem[Gendreau et al.(2012)]{gendreau12}Gendreau, K. C., Arzoumanian, Z., \& Okajima, T.\ 2012, Proc. SPIE, 8443, 13
\bibitem[Gladstone et al.(2007)]{gladstone07}Gladstone, J., Done, C., \& Gierli\'{n}ski, M.\ 2007, MNRAS, 378, 13
\bibitem[Gonz\'{a}lez Mart\'{i}nez-Pa\'{i}s et al.(2014)]{gonzalez14}Gonz\'{a}lez Mart\'{i}nez-Pa\'{i}s, I., Shahbaz, T., \& Casares Vel\'{a}zquez, J.\ 2014, Accretion Processes in Astrophysics (Cambridge: Cambridge Univ. Press)
\bibitem[Harrison et al.(2013)]{harrison13}Harrison, F. A., Craig, W. W., Christensen, F. E., et al.\ 2013, ApJ, 770, 103
\bibitem[Hasinger \& van der Klis(1989)]{hvp89}Hasinger, G., \& van der Klis, M. 1989, A\&A, 225, 79
\bibitem[Homan et al.(2010)]{homan10}Homan, J., van der Klis, M., Fridriksson, J. K., et al.\ 2010, ApJ, 719, 201
\bibitem[Homan et al.(2018)]{homan18}Homan, J. Steiner, J. F., Lin, D., et al.\ 2018, ApJ, 853, 157
\bibitem[Iaria et al.(2014)]{iaria14} Iaria, R., Di Salvo, T., Burderi, L., et al.\ 2014, A\&A, 561, 99
\bibitem[Iaria et al.(2016)]{iaria16}Iaria, R., Di Salvo, T., Del Santo, M., et al.\ 2016, A\&A, 596, 21 
\bibitem[Ibragimov \& Poutanen(2009)]{ip09}Ibragimov, A., \& Poutanen, J.\ 2009, MNRAS, 400, 492
\bibitem[King et al.(2016)]{king16}King, A. L., Tomsick, J. A., Miller, J. M., et al.\ 2016, ApJL, 819, L29
\bibitem[Kuulkers \& van der Klis(2000)]{kuulkers00}Kuulkers, E., van der Klis, M.\ 2000, A\&A, 356, L45
\bibitem[Kuulkers et al.(2002)]{kuulkers02}Kuulkers, E.\ 2002, A\&A, 383, L5
\bibitem[Kuulkers et al.(2003)]{kuulkers03}Kuulkers, E., den Hartog, P. R., in't Zand, J. J. M., et al.\ 2003, A\&A, 399, 663
\bibitem[Kuulkers et al.(2010)]{kuulkers10}Kuulkers, E., in't zand, J. J. M., Atteia, J.-L., et al.\ 2010, A\&A, 514, A65
\bibitem[Lamb et al.(1985)]{lamb85}Lamb, F. K., Shibazaki, N., Alpar, M. A., \& Shaham, J.\ 1985, Nature, 317, 681
\bibitem[Lamb et al.(2009)]{lamb09}Lamb, F. K., Boutloukos, S., Van Wassenhove, S., et al.\ 2009, ApJ, 706, 417 
\bibitem[Lin et al.(2007)]{lin07}Lin, D., Remillard, R. A., \& Homan, J.\ 2007, ApJ, 667, 1073
\bibitem[Lin et al(2010)]{lin10}Lin, D., Remillard, R. A., \& Homan, J.\ 2010, ApJ, 719, 1350
\bibitem[London, Taam, \& Howard(1986)]{london86}London, R. A., Taam, R. E., \& Howard, W. M.\ 1986, ApJ, 306, 170L
\bibitem[Ludlam et al.(2016)]{ludlam16} Ludlam, R. M., Miller, J. M., Cackett, E. M., et al.\ 2016, ApJ, 824, 37
\bibitem[Ludlam et al.(2017a)]{ludlam17a}Ludlam, R. M., Miller, J. M., Bachetti, M., et al.\ 2017a, ApJ, 836, 140
\bibitem[Ludlam et al.(2017b)]{ludlam17b}Ludlam, R. M., Miller, J. M., Cackett, E. M., Degenaar, N., \& Bostrom, A. C.\ 2017b, ApJ, 838, 79
\bibitem[Ludlam et al.(2017c)]{ludlam17c}Ludlam, R. M., Miller, J. M., Degenaar, N., et al.\  2017c, ApJ, 847, 135
\bibitem[Madej et al.(2010)]{madej10}Madej, O. K., Jonker, P. G., Fabian, A. C., et al.\ 2010, MNRAS, 407, L11
\bibitem[Madej et al.(2014)]{madej14}Madej, O. K., Garc\'{i}a, J., Jonker, P. G., et al.\ 2014, MNRAS, 442, 1157 
\bibitem[Makishima et al.(1983)]{makishima83}Makishima K., Mitsuda K., Inoue, H., et al.\ 1983, ApJ, 267, 310
\bibitem[Matranga et al.(2017)]{matranga17}Matranga, M., Di Salvo, T., Iaria, R., et al.\ 2017, A\&A, 600, A24
\bibitem[Markwardt et al.(1999)]{markwardt99}Markwardt, C. B., Strohmayer, T. E., \& Swank, J. H.\ 1999, ApJ, 512, L125
\bibitem[Mazzola et al.(2018)]{mazzola18}Mazzola, S. M., Iaria, R., Di Salvo, T., et al.\ 2018, Arxiv: 1811.10922
\bibitem[Merloni, Fabian, \& Ross(2000)]{merloni00}Merloni, A., Fabian, A. C., Ross, R. R.\ 2000, MNRAS, 313, 193
\bibitem[Miller et al.(2009)]{miller09}Miller, J. M., Cackett, E. M., \& Reis, R. C.\ 2009, ApJ, 707, L77
\bibitem[Miller et al.(2011)]{miller11}Miller, J. M., Maitra, D., Cackett, E. M., Bhattacharyya, S., \& Strohmayer, T. E.\ 2011, ApJ, 731, L7
\bibitem[Miller et al.(2013)]{miller13}Miller, J. M., Parker, M. L., Fuerst, F., et al.\ 2013, ApJ, 779, L2
\bibitem[Mitsuda et al.(1984)]{mitsuda84}Mitsuda, K., Inoue, H., Koyama, K., et al.\ 1984, PASJ, 36, 741
\bibitem[Mitsuda et al.(1989)]{mitsuda89}Mitsuda, K., Inoue, H., Nakamura, N., \& Tanaka, Y.\ 1989, PASJ, 41, 97
\bibitem[Mondal et al.(2018)]{mondal18}Mondal, A. S., Dewangan, G. C., Pahari, M, \& Raychaudhuri, B.\ 2018, MNRAS, 474, 2064
\bibitem[Narayan \& Yi(1994)]{narayan94}Narayan, R., \& Yi, I.\ 1994, ApJ, 428, L13
\bibitem[Nelemans et al.(2004)]{nelemans04}Nelemans, G., Jonker, P. G., Marsh, T. R., \& van der Klis, M.\ 2004, MNRAS, 348, L7
\bibitem[Oosterbroek et al.(2001)]{oosterbroek01}Oosterbroek, T., Barret, D., Guainazzi, M., \& Ford, E. C.\ 2001, A\&A, 366, 138
\bibitem[Papitto et al.(2009)]{papitto09}Papitto, A., Di Salvo, T., D'A\'{i}, A., et al.\ 2009, A\&A, 493, L39
\bibitem[Parmar et al.(1989)]{parmar89}Parmar, A. N., Stella, L., \& Giommi, P.\ 1989, A\&A, 222, 96
\bibitem[Petrucci et al.(2001)]{petrucci01}Petrucci, P. O., Merloni, A., Fabian, A., Haardt, F., \& Gallo, E.\ 2001, MNRAS, 328, 501
\bibitem[Pintore et al.(2015)]{pintore15}Pintore, F., Di Salvo, T., Bozzo, E., et al.\ 2015, MNRAS, 450, 2016
\bibitem[Piraino et al.(2012)]{piraino12}Piraino, S., Santangelo, A., Kaaret, P., et al.\ 2012, A\&A, 542, L27 
\bibitem[Popham \& Sunyaev(2001)]{ps01}Popham, R., \& Sunyaev, R.\ 2001, ApJ, 547, 355
\bibitem[Pritzl et al.(2001)]{pritzl01}Pritzl, B. J., Smith, H. A., Catelan, M., \& Sweigart, A. V.\ 2001, ApJ, 122, 2600
\bibitem[Ross \& Fabian(2005)]{ross05}Ross, R. R., \& Fabian, A. C. 2005, MNRAS, 358, 211
\bibitem[Sansom et al.(1993)]{sansom93}Sansom, A. E., Dotani, T., Asai, K., \& Lehto, H. J.\ 1993, MNRAS, 262, 429
\bibitem[Shahbaz et al.(2008)]{shahbaz08}Shahbaz, T., Watson, C. A., Zurita, C., Villaver, E., \& Hernandez-Peralta, H.\ 2008, PASP, 120, 848
\bibitem[Shimura \& Takahara(1995)]{shimura95}Shimura, T., Takahara, \& F.\ 1995, ApJ, 445, 780
\bibitem[Sibgatullin \& Sunyaev(2000)]{ss00}Sibgatullin, N. R. \& Sunyaev, R. A.\ 2000, Astronomy Letters, 26, 699
\bibitem[Sleator et al.(2016)]{sleator16}Sleator, C. C., Tomsick, J. A., King, A. L., et al.\ 2016, ApJ, 827, 134
\bibitem[Strohmayer et al.(2008)]{strohmayer08}Strohmayer, T. E., Markwardt, C. B., \& Kuulkers, E.\ 2008, ApJ, 672, L37
\bibitem[Swank et al.(1976)]{swank76}Swank, J. H., Becker, R. H., Pravdo, S. H., Saba, J. R., \& Serlemitsos, P. J.\ 1976, IAU Circ., 3010, 1
\bibitem[Swank et al.(1978)]{swank78}Swank J. H., Boldt E. A., Holt S. S., Serlemitsos P. J., \& Becker R. H.\ 1978, MNRAS, 182, 349
\bibitem[Sztajno et al.(1987)]{sztajno87}Sztajno, M., Fujimoto, M. Y., van Paradijs, J., et al.\ 1987 MNRAS, 226, 39
\bibitem[Tomsick et al.(2018)]{tomsick18}Tomsick, J. A., Parker, M. L., Garc\'{i}a, J. A., et al.\ 2018, ApJ, 855, 3
\bibitem[van der Klis(2005)]{van05}van der Klis, M. 2005, in NATO Advanced Science Institutes (ASI) Series B, Vol. 210, NATO Advanced Science Institutes (ASI) Series B, ed. A. Baykal, S. K. Yerli, S. C. Inam, \& S. Grebenev, 283
\bibitem[van den Eijnden et al.(2017)]{vandeneijnden17}van den Eijnden, J., Bagnoli, T., Degenaar, N., et al.\ 2017, MNRAS, 466, 98
\bibitem[van den Eijnden et al.(2018)]{vandeneijnden18}van den Eijnden, J., Degenaar, N., Pinto, C., et al.\ 2018, MNRAS, 475, 2027
\bibitem[Veledina, Poutanen, \& Vurm(2013)]{veledina13}Veledina A., Poutanen J., \& Vurm I.\ 2013, MNRAS, 430, 3196
\bibitem[Verner et al.(1996)]{vern}Verner, D. A., Ferland, G. J., Korista, K. T., \& Yakovlev, D. G.\ 1996, ApJ, 465, 487
\bibitem[Wijnands et al.(2017)]{wijnands17}Wijnands, R., Parikh, A. S., Altamirano, D., Homan, D., \& Degenaar, N.\ 2017, MNRAS, 472, 559
\bibitem[Wilkins(2018)]{wilkins18} Wilkins, D. R.\ 2018, MNRAS, 475, 748
\bibitem[Wilms et al.(2000)]{wilms}Wilms, J., Allen, A., \& McCray, R.\ 2000, ApJ, 542, 914
\bibitem[White et al.(1988)]{white88}White, N. E., Stella, L., \& Parmar, A. N.\ 1988, ApJ, 324, 363
\bibitem[Zdziarski et al.(1996)]{nthcomp1}Zdziarski, A. A.,  Johnson, W. N., \& Magdziarz, P.\ 1996, MNRAS, 283, 193
\bibitem[Zycki et al.(1999)]{nthcomp2}Zycki, P. T., Done, C., \& Smith, D. A.\ 1999, MNRAS 309, 561

\end{thebibliography}
\end{document}